\documentclass[
]{IEEEtran}

\usepackage{cite}
\usepackage{amsmath,amssymb}
\usepackage{graphicx}
\usepackage{color}
\usepackage{placeins}
\usepackage{float}
\usepackage{tabularx,colortbl}
\usepackage{ifthen}
\usepackage[acronym]{glossaries}
\usepackage{siunitx}
\usepackage{tabularx}
\usepackage{multirow}
\usepackage{booktabs}
\usepackage{flushend}
\usepackage[normalem]{ulem}
\usepackage{algorithm}
\usepackage{amsmath}
\usepackage{algpseudocode}
\usepackage{mathtools}
\usepackage{gensymb}
\usepackage{fancyhdr, lipsum}

 \algnewcommand{\Initialize}{%
  \State \textbf{Initialize:}
}

\newacronym{crlb}{CRLB}{Cramer-Rao lower bound}
\newacronym{qls}{QLS}{quadratic-least squares}
\newacronym{sdr}{SDR}{software-defined radio}
\newacronym{bpsk}{BPSK}{binary phase shift keying}
\newacronym{isac}{ISAC}{integrated sensing and communication}
\newacronym{snr}{SNR}{signal-to-noise ratio}
\newacronym{ber}{BER}{bit-error rate}
\newacronym{if}{IF}{intermediate frequency}
\newacronym{awg}{AWG}{arbitrary waveform generator}
\newacronym{v2x}{V2X}{vehicle-to-everything}
\newacronym{iot}{IoT}{Internet-of-Things}
\newacronym{ofdm}{OFDM}{orthogonal frequency division multiplexing}
\newacronym{mimo}{MIMO}{multiple--input multiple--output}
\newacronym{lo}{LO}{local oscillator}
\newacronym{prbs}{PRBS}{pseudo-random bit sequence}
\newacronym{fom}{FOM}{figure-of-merit}
\newacronym{lfm}{LFM}{linear frequency modulated}
\newacronym{cdma}{CDMA}{code divison multiplexing access}
\newacronym{miso}{MISO}{multiple input single output}
\newacronym{csi}{CSI}{channel state information}
\newacronym{pam}{PAM}{Phase and Amplitude Modulated}
\newacronym{qpsk}{QPSK}{Quadrature Phase Shift Keying}
\newacronym{psk}{PSK}{phase shift keying}
\newacronym{ser}{SER}{Symbol Error Rate}
\newacronym{svd}{SVD}{Singular Value Decomposition}
\newacronym{siso}{SISO}{Single Input Single Output}
\newacronym{ula}{ULA}{Uniform Linear Array}
\newacronym{pll}{PLL}{Phased-Locked Loop}
\newacronym{evm}{EVM}{Error Vector Magnitude}
\newacronym{ml}{MLE}{Maximum Likelihood Estimator}
\newacronym{los}{LOS}{Line-of-Sight}
\newacronym{cfo}{CFO}{Carrier Frequency Offset}
\newacronym{an}{AN}{Artificial Noise}
\newacronym{sinr}{SINR}{Signal-to-Interference and Noise Ratio}

\begin{document}
\title{Secure Wireless Communication Using Coherent Distributed Transmission and Spatial Signal Decomposition}
	
		\author{Anton Schlegel,~\IEEEmembership{Graduate Student Member,~IEEE}, Jason M. Merlo,~\IEEEmembership{Member,~IEEE}, Samuel Wagner,\\John B. Lancaster, and Jeffrey A. Nanzer,~\IEEEmembership{Senior Member,~IEEE}
		\thanks{Manuscript received 2025.}
		\thanks{This work was supported in part by the Google Research Scholar program, in part by the National Science Foundation under grant \#2225337 and \#2534114, and in part by the LLNL LDRD Program under Project No. 22-ER-035 and 25-ER-040. Release number: LLNL-JRNL-2013373-DRAFT.}
		\thanks{A. Schlegel, J. M. Merlo and J. A. Nanzer are with the Department of Electrical and Computer Engineering, Michigan State University, East Lansing, MI 48824 USA (e-mail: schleg19@msu.edu; merlojas@msu.edu; nanzer@msu.edu).}
		\thanks{S. Wagner and J. B. Lancaster are with the Lawrence Livermore National Laboratory, Livermore, CA 94550 USA.}
		}
		
	\maketitle
	\begin{abstract}
	
	We present a new approach to secure wireless communications using coherent distributed transmission of signals that are spatially decomposed between a two-element distributed antenna array. High-accuracy distributed coordination of microwave wireless systems supports the ability to transmit different parts of a signal from separate transmitters such that they combine coherently at a designated destination. In this paper we explore this concept using a two-element coherent distributed phased array where each of the two transmitters sends a separate component of a communication signal where each symbol is decomposed into a sum of two pseudo-random signal vectors, the coherent summation of which yields the intended symbol. By directing the transmission to an intended receiver using distributed beamforming, the summation of the two vector components is largely confined to a spatial region at the destination receiver. We implement the technique in a $\SI{50}{\lambda}$ array operating at \SI{3}{\giga\hertz}. We evaluate the symbol error ratio (SER) in two-dimensional space through simulation and measurement, showing that the approach yields a spatially-confined secure region where the information is recoverable (i.e., the received signal has low SER), and outside of which the information is unrecoverable (high SER). The proposed system is also compared against a traditional beamforming system where each node sends the same data. We validate experimentally that our approach achieves a low SER of 0.0082 at broadside and a SER above 0.25 at all other locations compared to a traditional beamforming approach that achieves a SER of 0 at all locations measured.

	\end{abstract}
	
\makeatletter
\def\endthebibliography{%
  \def\@noitemerr{\@latex@warning{Empty `thebibliography' environment}}%
  \endlist
}
\makeatother

\section{Introduction}

Rapid advancements in wireless communications technology has led to an increasing number of wireless connections in existing systems like 5G and automotive V2X networking, and will lead to a further increase in emerging systems like 6G and beyond~\cite{9598915, 7399671, 7414384, 8879484, raya2005security, Dang:2020aa, 9509294, 10820534}. A greater number of wireless connections, whether used for communications, sensing, or both, entails high susceptibility to security issues such as intentional or unintentional interference, eavesdropping, and man-in-the-middle (MITM) attacks, among other concerns~\cite{yeh2016security, Zhang:2025aa, 10949709}. Eavesdropping and MITM are particularly concerning because the malicious actor captures transmitted signals passively, making it challenging to detect the presence of the receiver and to apply additional mitigation techniques. By their nature, wireless systems transmit energy in a broad range of directions, making it challenging to safeguard transmitted information. As the number of wireless connections in emerging networks increases, the ability to safeguard the transmission of information will become more important and also more challenging.

Securing transferred information has traditionally focused on methods like cryptography, where the information is encoded and transferred over an unsecured wireless channel~\cite{Najm:2025aa, 10829860, cryptography1010004}. Thus, while the data itself is encrypted, it can nevertheless be intercepted in the transmission process, and may be vulnerable to decryption later, which has led to interest in physical layer approaches that mitigate the transmission of information to unwanted locations. Beamforming provides spatial filtering, but only reduces the SNR at directions outside of the mainbeam \cite{9849675}. Even with highly directive phased arrays in future high-frequency networks like 6G, sidelobe structure nonetheless imparts a transmission of information in many directions away from the intended receiver, which can be detected by a malicious receiver that is nearby or has higher sensitivity to overcome the lower signal-to-noise ratio (SNR) due to the lower sidelobe gain. Another approach utilizes injecting artificial noise into the transmitted waveforms to increase the \gls{sinr} at the eavesdropper~\cite{4543070, 1558439, 9968163, 10153696}, however this still serves only to reduce the signal level, and do not affect the underlying information. Approaches that change the transmitted information as a function of space, referred to as directional modulation, are thus of interest~\cite{10161710, 10236567, 8333706, 10217146, 10295410}. In this approach, the physical aperture is generally modulated during transmission in order to add additional distorting modulation to the transmitted signal, thereby obscuring the actual information and making decryption more difficult outside of an intended secure region. Notably, such physical layer approaches can be used on conjunction with traditional cryptographic techniques. A similar approach is to use multiple-input, multiple-output (MIMO) approaches where separate parts of the information are transmitted from separate antennas, which can be implemented in phased arrays~\cite{8855017}.

\begin{figure*}[t!]
\centering
	\includegraphics[width=0.8\textwidth]{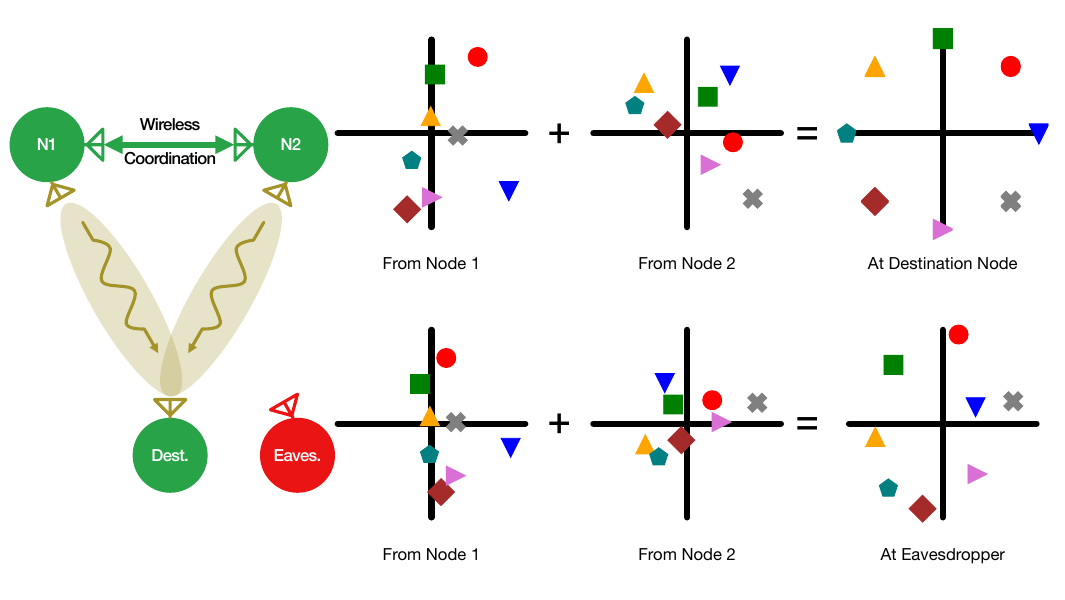}
	\caption{Proposed concept for distributed transmission of spatially decomposed signals. Node 1 and Node 2 generate pseudo-random sequences that sum correctly only at the intended destination or receiver.}
	\label{fig:secure_comms_example}
\end{figure*}

The approaches above are monolithic in that they are implemented on a single platform system, however the increasing diversity of wireless systems in emerging networks provides an opportunity to use distributed techniques for an additional layer of wireless security. In particular, implementing techniques leveraging distributed coherence between platforms provides unique capabilities for new security techniques. Coordination between the separate systems is challenging, however, since the signal transmission must be aligned at the wavelength level, necessitating high accuracy coordination of time (synchronization), frequency (syntonization), and phase (which is obtained through localization). Distributed phased arrays are wireless networks with appreciable electrical separation between platforms~\cite{1088883} that are wirelessly coordination to support phase-coherent beamforming~\cite{7803582, 9492307}. There are several benefits to using distributed arrays over single monolithic platforms, including: the distribution of information among several transceivers removes any single point of failure in the system, making it robust to interference; the nodes can move freely from one another, allowing dynamic array factor patterns to be generated, thus changing the power distribution; because the nodes are not physically connected, additional nodes can easily be added or removed from the array. The main challenges with distributed arrays revolve around the coordination requirements, which necessitate wavelength-level alignment of the electrical states in a fast and reliable fashion. Extensive prior work has, however, been conducted to create wireless technologies that support distributed coherence at microwave frequencies, with time, phase, and frequency alignment implemented in various experimental systems and used to demonstrate distributed beamforming~\cite{7218555, 9210120, 9173801, 9994246, 10443654, 10522855, 10880359, 11026108, merlo2025collab}.

In this paper we describe and experimentally demonstrate a new coherent distributed array approach that transmits portions of a spatially decomposed communication signal from two separate transmitters coordinated at the wavelength level. The transmitted vector signal information is decomposed into a summation of two pseudo-random subvectors, the superposition of which yields the correct data symbol. The array transmits the two subvectors on a carrier frequency, which is beamsteered to a desired receiver. Because of the separate path lengths encountered by the signal at angles away from the intended receiver, the superposition of the two subvectors does not equate to the intended symbol, thereby adding an layer of security on the transmitted data (see Fig.~\ref{fig:secure_comms_example}). We experimentally demonstrate the concept in a two node system with antennas separated by $\SI{50}{\lambda}$. Each node contains a \gls{sdr} that performs beamforming and coordination between the two nodes using the syntonization approach from~\cite{9210120} and the synchronization approach from~\cite{9994246}. The two node setup is able to achieve a \gls{ser} of 0.0082 at broadside and \gls{ser} above 0.25 for the measured eavesdropper positions.

\section{Distributed Array Coordination}

To support coherent transmission of the carxrier signals onto which the data is modulated, the electrical states of the transmitting systems must be appropriately aligned. For microwave transmission, time alignment on the order of picoseconds is necessary, while relative positioning on the order of centimeters or lower is needed for appropriate phase alignment. The relative clock frequencies must also be syntonized sufficiently often to ensure that the transmitters remain phase coherent over appreciable durations of the signals between updates. Precise thresholds depend on the carrier frequency and desired beamforming performance~\cite{9492307}, and various technologies have been developed by our group and others to support sufficient wireless coordination for distributed beamforming at microwave frequencies. In this section we review the approaches to time synchronization and frequency syntonization. In this work the nodes are not moved relative to one another; in the case where relative motion is imparted, localization can be implemented using the same data as is used in the synchronization approach shown below~\cite{11026108}.

\subsection{Synchronization and Localization Based on Two-Way Time Transfer}

Synchronization of the transmitted signals ensures that the two signals arrive at the intended destination with sufficient temporal overlap to ensure a that the summation of the two data subvectors accurately reconstructs the intended symbol. Localization is needed to estimate the relative phase offset to implement the beamsteering operation. Both can be obtained through a two-way time transfer approach as described in~\cite{9994246,11026108}. In this method, a given node $m$ transmits a signal to a neighboring node $n$, which subsequently retransmits a signal back to node $m$. This two-way exchange of signals yields four timestamps: $t_{\mathrm{RX}m}$ and $t_{\mathrm{TX}m}$, which are the receive and transmit times on the first node respectively, and $t_{\mathrm{RX}n}$ and $t_{\mathrm{RX}n}$, which are the receive time and transmit time on the secondary node. The differences between the transmitted and received signals yields the apparent times of flight between the nodes, given by
\begin{equation}
\tau_{m,n} = (t_{\mathrm{RX}m} - t_{\mathrm{TX}n})
\end{equation}
which includes the clock error in each node. %
Once the four timestamps are estimated, the offsets in the clocks between the nodes can be calculated by
\begin{equation}
\Delta t = \frac{1}{2}\left[(t_{\mathrm{RX}m} - t_{\mathrm{TX}n}) - (t_{\mathrm{RX}n} - t_{\mathrm{TX}m})\right]
\end{equation}
Once the timing offset $\Delta t$ is estimated, node $m$ can adjust its local clock edge or adjust the transmission time of the waveform to compensate for the timing difference.
The transmit timestamps are known to the precision of the clock times, which for the \gls{sdr}s used in this work (Ettus USRP X310) is accurately known to below a picosecond. The reception times $t_{\mathrm{RX}i}$ thus must be estimated with errors on the order of picoseconds to ensure accurate synchronization. This is accomplished using a spectrally-sparse, two-tone signal that yields a near-optimal estimate of the reception time~\cite{9057428}. This approach has been demonstrated to achieve a timing precision of \SI{2.26}{\pico \second} with a \SI{40}{\mega \hertz} two-tone waveform \cite{9994246}. 

Once the four timestamps are estimated, the total time of flight $T_{m,n}$ between the two systems can also be estimated, from which the relative distance between the two nodes can be determined after calibrating for the static signal processing latency and RF front end transmission line delays $\delta T_{m,n}$ in node $n$. The calibrated time of flight is given by
\begin{equation}
T_{m,n} = \frac{1}{2}\left[(t_{\mathrm{RX}m} - t_{\mathrm{TX}n}) + (t_{\mathrm{RX}n} - t_{\mathrm{TX}m})\right] - \delta T_{m,n}
\end{equation}
From the calibrated time of flight the relative distance can be estimated by multiplying by the speed of light, $d = cT$. The relative phase needed for beamsteering to a desired location can then be determined by estimating the distance relative to the carrier frequency wavelength $\lambda = \frac{c}{f}$. This approach has shown demonstrated localization accuracies on the order of millimeters~\cite{11026108}.

\subsection{Frequency-Transfer}

\begin{figure}[t!]
	\includegraphics{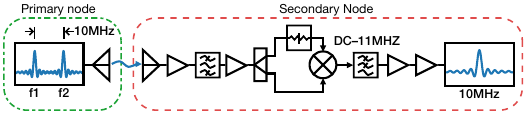}
	\caption{Frequency locking block diagram to transmit a frequency reference from a primary node to a secondary node. The primary node generates the two-tone waveform that is self-mixed by the secondary node to produce the frequency reference.}
	\label{freq_block_diagram}
\end{figure}

Communications signals are susceptible to phase and frequency offset since they can alter the demodulated symbols, While monolithic systems can rely on receive-side techniques like Costas loops, this cannot correct for frequency differences between two simultaneous transmitters, which must therefore be corrected wirelessly between the transmit nodes. In this work we use a wireless frequency transfer technique demonstrated in~\cite{9210120, 7218555}, and shown in Fig. \ref{freq_block_diagram}. The primary node generates a two-tone waveform, the frequency separation of which is the desired reference frequency, in this work \SI{10}{\mega \hertz}. The secondary node receives the two-tone signals which is then input to a self-mixing circuit that demodulates the reference frequency signal. The self-mixing circuit includes a mixer and an attenuator on the LO path of the mixer to ensure equal amplitude of the two input signals. A filter is used to remove any higher-frequency signals, and then amplified. The demodulated frequency references is then input to a phase-locked loop on the secondary node, enabling a wireless frequency lock.

\subsection{Node Calibration}
Calibration can be performed using various techniques, the goal of which is to estimate and correct for static phase offsets in the system. The techniques described above are aimed mainly at correcting dynamic phase errors encountered in the environment and frequency errors from clock drift. System-level static phase errors may be estimated in-situ using loop-backs from the antenna input terminals to estimated the phase delay on each transceiver, or they may be characterized a-priori if the phase delays are expected to be repeatable. In a communications system, the receiving node may also be used to provide information on the relative phase alignment of the transmitters. Since the objective of this paper is to demonstrate the feasibility of the spatially decomposed signal transmission approach, and because of the extensive literature on wireless system coordination, we use the receiving node to correct for static phase offsets in this work, after which the wireless coordination between the two nodes is relied upon for coherent distributed transmission.
The calibration is performed with a training pulse sent from each node. The calibration is performed at the beginning with all following pulses using the same calibration. Node 1 transmits a \SI{160}{\mega \hertz}, \SI{5}{\micro \second} long up-chirp \gls{lfm} waveform and node 2 transmits a \SI{160}{\mega \hertz}, \SI{5}{\micro \second} long down-chirp \gls{lfm}. The two waveforms are received combined and processed via matched filters. Both the up-chirp and down-chirp are matched to their respective baseband signals. Node 2's static phase and amplitude values are compared against Node 1's and corrected using 
\begin{equation}
	\frac{B_{1}}{B_{2}}e^{j\theta_{2} - j\theta_{1}}\delta(t - \tau_{n})
\end{equation}
where $B_i$ are the amplitudes of the signals, the phases $\theta_{i}$ are the static phases between each transmitter and the receiver, and $\tau_{n}$ is the relative delay between the nodes. 

\section{Communication Signal Decomposition}

We implement the proposed spatial decomposition approach using 8-\gls{psk} modulation as a proof-of-concept. The approach is shown in Fig. \ref{fig:secure_comms_example} and leverages the differences in physical and waveform dynamics to distort the data at undesired angles. Because the waveforms are only coordinated to align appropriately at the intended receiver, relative phase differences to other spatial locations will cause a summation that results in an incorrect symbol value.

\begin{figure}[t!]
\centering
\includegraphics[width=0.8\columnwidth]{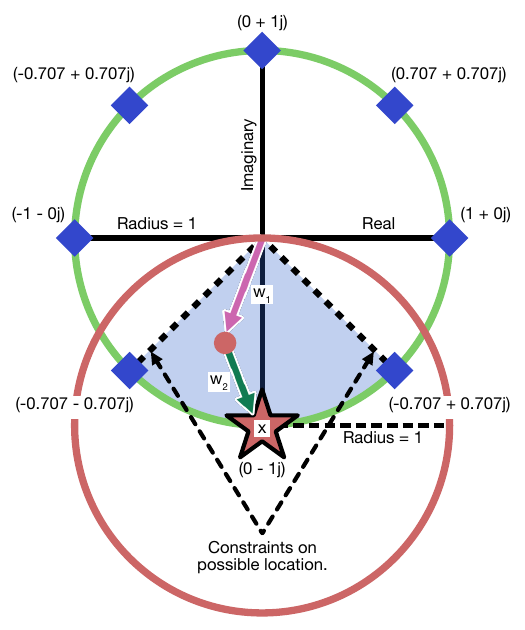}
\caption{Demonstration of the proposed secure algorithm. Node 1 transmits $\text{w}_1$, pink vector, and node 2 transmits $\text{w}_2$, the green vector. The vectors sum coherently at the receiver which is x, the red star.}
\label{fig:points}
\end{figure}

For the two node array, the data generation approach is shown in Fig.~\ref{fig:points}. 
The intended data point is decomposed into two pseudo-random data points that sum to the original point. The red star indicates the location of the desired symbol, which is given by the vector $\vec{x}$, and can be writtenin terms of the superposition of two subvectors by 
\begin{equation}
	\vec{x}[n] = \vec{w_1}[n] + \vec{w_2}[n]
\end{equation}
where $n$ is the data index and $\vec{w_i}$ are the two pseudo-random data subvectors. Each element in $\vec{x}$ is comprised of a complex data symbol, $\vec{x}[n] = \sqrt{E_s}e^{j \xi}$, where $E_{s}$ is the signal energy, and $\xi$ is the phase of modulated signal. 

We assume that each node transmits a normalized amplitude of unity. This generates a validity region, wherein the two subvectors must reside in order to appropriately reconstruct the desired signal vector. 
Since each transmitter is limited to unit amplitude, the valid region is characterized by the overlap between the unit circles around the origin and the desired data point. However, operating near the edges of this overlap region caused increased errors in experiment, thus we further restrict the valid region to be within an angle of $\pm 45\degree$ around the designated received point. 
The initial subvector data point, $\vec{w_1}[n]$, is randomly chosen within the valid region and is shown as the pink arrow in Fig.~\ref{fig:points}. The second subvector data point, $\vec{w_2}[n]$ is then chosen by
\begin{equation} \label{eqn:second_pt}
	\vec{w_2}[n] = \vec{x}[n] - \vec{w_1}[n].
\end{equation}

Note that the above formulation is valid only when the propagation phases are appropriately corrected, i.e., the static phase offsets and beamforming phases are implemented. At the intended receiver, if the wireless coordination is implemented, these phases are corrected and the superposition of the two subvectors yields the correct data vector. However, at other angles the beamforming phase is different, causing the subvectors to have additional phase delays added (which may be leading or lagging phase delays), which in turn generates a different value upon superposition, corrupting the transmitted information.

\section{Distributed Data Transmission Simulation}

We simulate a two node distributed array system beamforming to a receiver in the near-field of the array but far-field to each individual antenna. The array is assumed to be perfectly coordinated, and no noise is implemented in the simulation; these assumptions ensure that any symbol errors are generated entirely by the signal decomposition method and the relative phase differences imparted on the subvectors at different locations in space, thereby providing an analysis on the security benefits of the approach in isolation.
The simulation parameters are given in Table~\ref{table:sim_distances}. The minimum distance between the transmitters and the receiver was $\SI{5}{\lambda}$, thus any near-field propagation effects could be neglected and thus only the amplitude dependence relative to $1/r$ was included.
The locations of the transmitters and receiver are given in Table~\ref{table:2}. 
The signal parameters are given in in Table~\ref{table:3}. We use 10 times oversampling of the data to ensure that the data is not under-discretized, and we evaluate the performance by calculating the SER in two-dimensional space in spatial increments of $1\lambda$. 
To demodulate the data, a \gls{ml} that calculates the Euclidean distance between the received point and the estimated point is used. An \gls{prbs} with a length of 300 bits was modulated with 8-\gls{psk} and then split into the two subvector data streams, each transmitted from the two antenna locations. 
100 Monte Carlo simulations were run with different randomized signal decompositions.
Phase, amplitude, and time compensation are added to the communication waveforms to ensure that they sum coherently at the intended receiver. 
The simulations were parameterized for \SI{1}{\giga \hertz}, \SI{2}{\giga \hertz}, and \SI{3}{\giga \hertz}. For all simulations, isotropic radiators were assumed.

\begin{table}[t!]
\begin{center}
\caption{\label{table:sim_distances}The simulation parameters for the symbol rate analysis simulation.}
\begin{tabular}{p{0.5in} p{0.6in} p{0.45in}  p{0.45in} p{0.6in}} \toprule
Array Baseline & Receiver Distance &Far Field & Simulation Width  & Simulation Height\\
\midrule
$50 \lambda$ & $50 \lambda$ & $5000 \lambda$  & $100 \lambda$ & $95 \lambda$\\
 \bottomrule
\end{tabular}
\end{center}
\end{table}

\begin{table}[t!]
\begin{center}
\caption{\label{table:2}The location of the transmitters and receivers.}
\begin{tabularx}{\columnwidth}{*2{X}}
\toprule
Transceiver & Location (m, m) \\ 
\midrule
Tx0 & ($-25 \lambda$, 0) \\
Tx1 & ($25 \lambda$, 0)\\
Receiver & (0, $50 \lambda$) \\
Eavesdropper & (Moves through search space) \\
\bottomrule
\end{tabularx}
\end{center}	
\end{table}

\begin{table}[t!]
\begin{center}
\caption{\label{table:3}Data rate and sample rate used for the simulation.}
\begin{tabularx}{\columnwidth}{*2{X}}
\toprule
 & Data Rate \\ 
\midrule
Symbol Rate & \SI{20}{\mega Sym/\second} \\
Sample Rate & \SI{200}{\mega Sa/\second}\\
\bottomrule
\end{tabularx}
\end{center}	
\end{table}

\begin{figure*}[t!]
\centering
\includegraphics[width=0.75\textwidth]{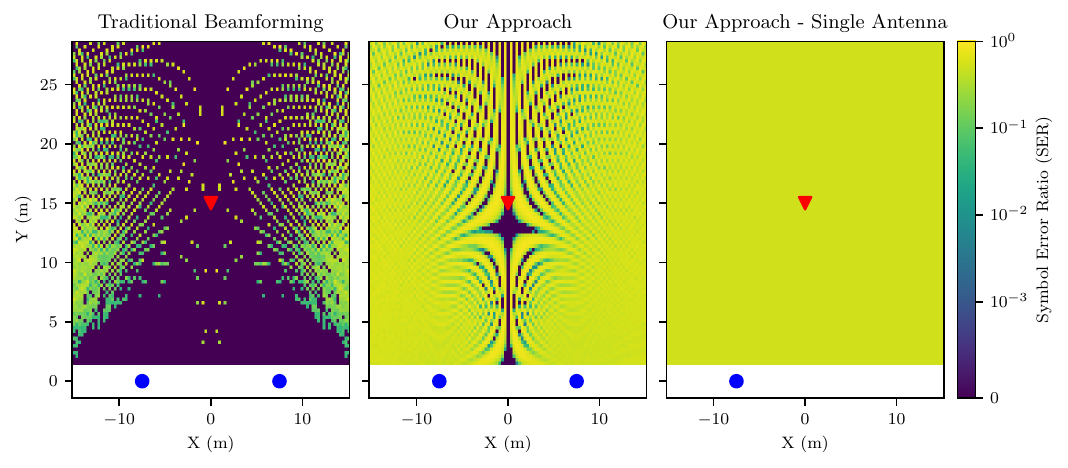}

(a)

\includegraphics[width=0.75\textwidth]{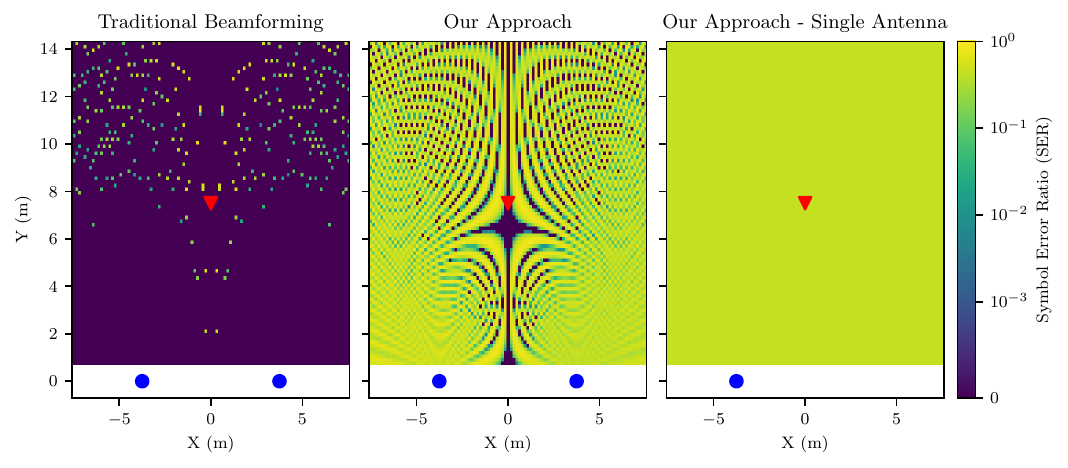}

(b)

\includegraphics[width=0.75\textwidth]{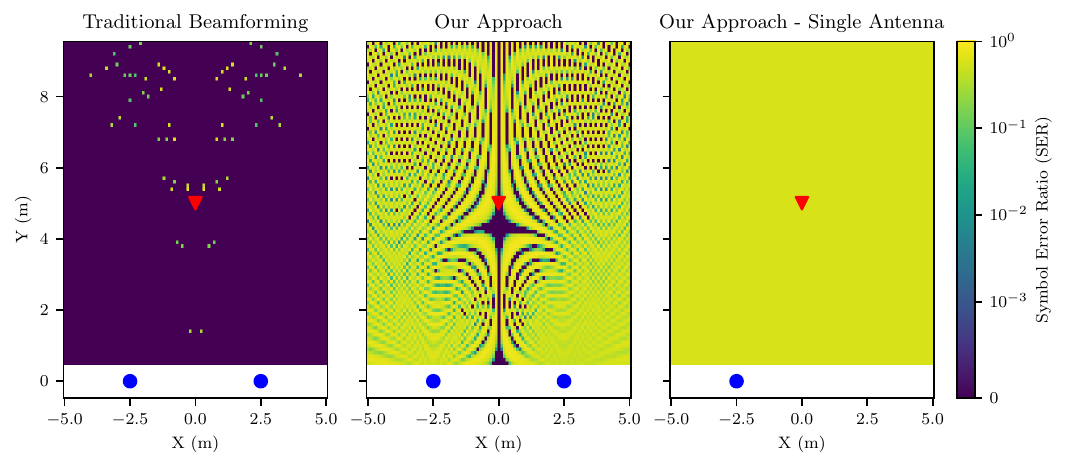}

(c)

\caption{Simulation results for (a) \SI{1}{\giga \hertz}, (b) \SI{2}{\giga \hertz}, and (c) \SI{3}{\giga \hertz}. The transmitters are indicated by blue circles, the receiver by the red triangle.}
\label{fig:sim_1ghz}
\end{figure*}

Three different types of transmission systems were evaluated. The first assumed a distributed antenna array with perfect coordination where both antennas transmitted the intended signal $\vec{x}$, which is referred to as the {traditional beamforming} approach. The second type is the proposed secure transmission approach using spatial signal decomposition. The third is to demonstrate that a single subvector component cannot be used to recover the data, and is the transmission from one antenna of only the subvector $\vec{w_1}$.
Fig.~\ref{fig:sim_1ghz} shows the simulated results for \SI{1}{\giga \hertz}, \SI{2}{\giga \hertz}, and \SI{3}{\giga \hertz}. The blue circles are the locations of the transmitters and the red triangle is the location of the receiver. 
High \gls{ser} is given in yellow and low \gls{ser} is given in purple. It is clear that the traditional beamforming approach results in low SER and thus recoverable data over the majority of the region. The proposed secure approach, however, significantly reduces the areas where the data is recoverable, resulting in only a few small regions of low SER. The single antenna case shows high SER everywhere, confirming that the subvector cannot be used to reconstruct the original data.
Table \ref{Table:ser_sim} gives the minimum and maximum \gls{ser} along with the percentage of the simulations with a \gls{ser} above 0.1. 
In the traditional beamforming case, \SI{2}{\giga \hertz} and \SI{3}{\giga \hertz} have \gls{ser} above 0.1 over only 2\% of the region, while for \SI{1}{\giga \hertz}  it is 20\%. 
In using the proposed secure approach, at least 78\% of the area achieves at \gls{ser} of 0.1 for all frequencies simulated.

\begin{table}
\caption{Information from Simulations}
\label{Table:ser_sim}
\begin{center}

\begin{tabular}{p{0.75in} p{0.5in} p{0.5in} p{1in}}
\toprule
Type & Min \gls{ser} & Max \gls{ser} & Percentage of Area Above \gls{ser} 0.1 \\
\midrule
\multicolumn{4}{c}{1 GHz} \\
\midrule
 Traditional & 0 & 0.79 & 20.61\\
 Secure & 0 & 0.88 & 88.08 \\
Single Antenna & 0.51 & 0.51 & 100\\
\midrule
\multicolumn{4}{c}{2 GHz} \\
\midrule
Traditional & 0 & 0.69 & 1.65 \\ 
Secure & 0 & 0.89 & 78.2 \\
Single Antenna & 0.44 & 0.44 & 100\\
\midrule
\multicolumn{4}{c}{3 GHz} \\
\midrule
Traditional & 0 & 0.75 & 0.63 \\ 
Secure & 0 & 0.91 & 78.85 \\
Single Antenna & 0.57 & 0.57 & 100\\
\bottomrule
\end{tabular}
\end{center}	
\end{table}

\section{System and Experimental Design}

\subsection{Hardware Design}
\begin{figure*}[t!]
	\centering
	\includegraphics[width=0.85\textwidth]{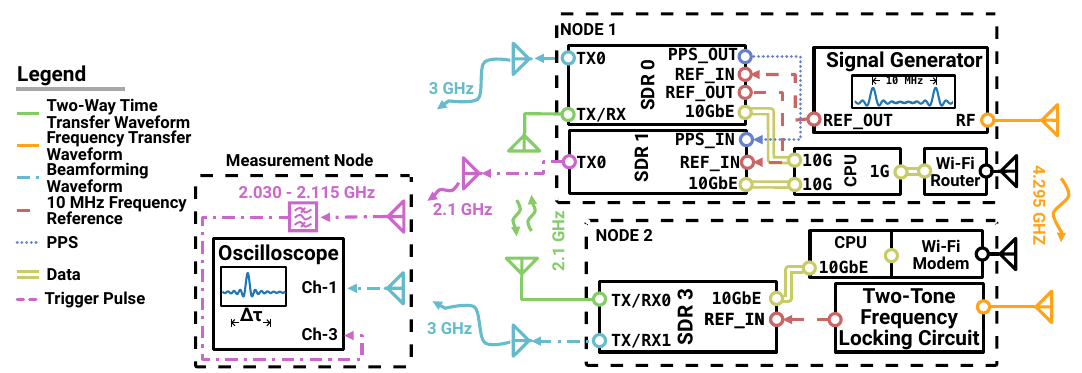}
	\caption{Block diagram of the experimental setup.}
	\label{block_diagram}
\end{figure*}

The block diagram of the hardware is shown in Fig. \ref{block_diagram}. Each node used Ettus X310 \gls{sdr}s, which have an analog bandwidth of \SI{160}{\mega \hertz} and a sample rate of \SI{200}{\mega Sa/\second}. Node 1 used two \gls{sdr}s, one to transmit and receive the time transfer waveform to Node 2, and the other for beamforming. The second \gls{sdr} sent a trigger waveform to the receiving node (an oscilloscope) to collect the calibration and beamforming waveforms. The beamforming and time transfer signals were transferred through log-periodic antennas (L-COM HG2458), the trigger signal was transferred through a different log-periodic antenna (L-COM HG7210LP), and the frequency transfer antenna was a standard gain horn antenna (Narda Model 643). Node 2 had a single \gls{sdr} which performed the same functions as the first \gls{sdr} on Node 1 and used the same style of antennas for its beamforming and time-transfer systems. 
The wireless frequency locking used a signal generator (Keysight E8267D) to generate the frequency reference to send from Node 1 to Node 2 using the standard gain horn antenna. 
Each node had a computer which controlled the \gls{sdr}s and which communicated via Wi-Fi for the beamforming scheduling between the nodes. 
An \SI{80}{\giga Sa/\second} (MSO-X 92004A) oscilloscope sampled at \SI{20}{\giga Sa/\second} was used as the receiver for the experiment and used a log-periodic antenna (L-COM HG2458) for receiving the beamformed signals and a log-periodic antenna (L-COM HG7210LP) along with a \SI{2.1}{\giga \hertz} cavity filter to capture the trigger signal. 
The received beamforming signals were processed offline.

\subsection{Software Design}
The system used GNU Radio in a distributed processing setup to control the hardware. The main components were a node controller, a time transfer controller, and a beamforming controller. The node controller sends and receives signals from the \gls{sdr}. The time-transfer controller runs the two-way time transfer algorithm described above and applies the timing correction. The beamforming controller generates the data to be sent from each of the nodes and also the trigger signal to the receiver. The computer on Node 1 ran the time-transfer, the node controller, and the beamforming controller software while the computer on Node 2 ran only a node controller. The beamforming controller on Node 1 generated the subvector data streams and transferred the second subvector data stream to Node 2 via Wi-Fi.

\subsection{Experimental Setup}

\begin{figure*}[t!]
	\centering
	\includegraphics[width=0.8\textwidth]{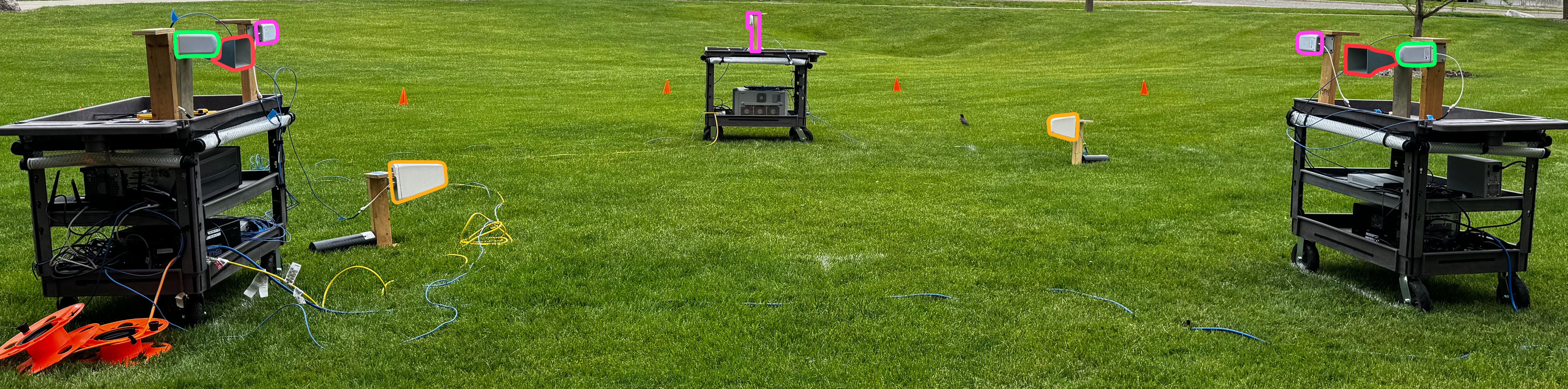}
	\caption{Experimental setup. Node 1 and Node 2 are in the foreground with the receiving node in the background. The orange antennas are for triggering the oscilloscope. The pink antennas are the beamforming antennas, the green for time-transfer, and the red for frequency transfer.}
	\label{node_full}
\end{figure*}

The outdoor setup is shown in Fig. \ref{node_full}. The carrier frequency used for data transmission to the receiver was \SI{3}{\giga \hertz}. The two-way time transfer approach used a carrier frequency of \SI{2.1}{\giga \hertz} and the frequency transfer used a carrier frequency of \SI{4.295}{\giga \hertz} for the first tone and \SI{4.305}{\giga \hertz} for the second tone. These frequencies were chosen to minimize self-interference within the array along with minimizing external interference. To trigger the oscilloscope to begin capturing the incoming data, a \SI{500}{\nano \s} wireless trigger pulse was sent on a \SI{2.1}{\giga \hertz} carrier frequency \SI{500}{\nano \s} before the transmission of the data. The oscilloscope captured \SI{140}{\kilo points} or \SI{7}{\micro \s} of data on each measurement. 
The trigger pulse and data were scheduled to transmit when the time transfer waveforms were not transmitting. The sample rate for each of the \gls{sdr}s on the nodes was \SI{200}{\mega Sa/\second}. The oscilloscope had a sample rate of \SI{20}{\giga Sa/\second}. Node 1 transmitted an up-chirp \gls{lfm} and node 2 transmitted a down-chirp \gls{lfm} before the communication waveforms were sent. The first up-chirp and down-chirp were used for calibrating the subsequent waveforms. The up-chirp was a \SI{160}{\mega \hertz} \gls{lfm} with the down-chirp being a \SI{160}{\mega \hertz} \gls{lfm}. Each calibration pulse was \SI{5}{\micro \s} long. On each subsequent transmit, the up-chirp and down-chirp were stored and used to characterize the phase, and amplitude stability of the subsequent transmitted pulses. Each node transmitted a subvector component of a decomposed 8-\gls{psk} waveform. 112 symbols were transmitted which consisted of permutations between the 8 constellation points in order to transmit the domain of possible phase transitions between constellation points. The communication waveform sent from each node had a symbol rate of \SI{20}{\mega Sym/\s} with a transmission time of \SI{5.6}{\micro \s}. 

The starting positions for the two nodes and receiver in the spectrally sparse array are shown in Table \ref{Table:positions}. The experimental setup is shown in Fig.~\ref{node_full}. The origin of the system was chosen to be the center between the two nodes. 
The system was calibrated to broadside and the receiver was moved in \SI{1}{\meter} steps to measure the \gls{ser} at off-broadside angles (eavesdropper locations). The locations of the receiver at each of the six positions are shown in Table \ref{Table:movement}. The trigger antenna placement was chosen to minimize its movement to ensure consistent power levels for the trigger threshold on the scope.

At each location, 25 measurements were conducted and then the receiver moved approximately \SI{1}{\meter} to the right towards node 2. 
The received data was demodulated with a hard decision demodulator. Because of coordination, multipath, and radio signal interference, we limited the search for \gls{cfo} compensation to $\pm$ \SI{50}{\kilo \hertz}. 

\begin{table}[t!]
\begin{center}
\caption{\label{Table:positions}The location of the transmitters and receivers.}
\begin{tabularx}{\columnwidth}{*2{X}}
\toprule
Transceiver & Location (m, m) \\ 
\midrule
Node 0 & (-2.56, 0) \\
Node 1 & (2.49 , 0)\\
Receiver & (0, 5.34) \\
\bottomrule
\end{tabularx}
\end{center}	
\end{table}

\begin{table}[t!]
\begin{center}
\caption{\label{Table:movement}The location of the receiver.}
\begin{tabularx}{\columnwidth}{*6{X}}
\toprule
Pos. 1 (m, m) & Pos. 2 (m, m) & Pos. 3 (m, m) & Pos. 4 (m, m) & Pos 5 (m, m) & Pos. 6 (m, m) \\ 
\midrule
(0, 5.34) &(0.98, 5.34) & (2.02, 5.34) & (2.98, 5.34) & (3.97, 5.34) & (5.02, 5.34)\\
\bottomrule
\end{tabularx}
\end{center}	
\end{table}

\begin{figure*}[t!]
	\centering
	\includegraphics[width=0.75\textwidth]{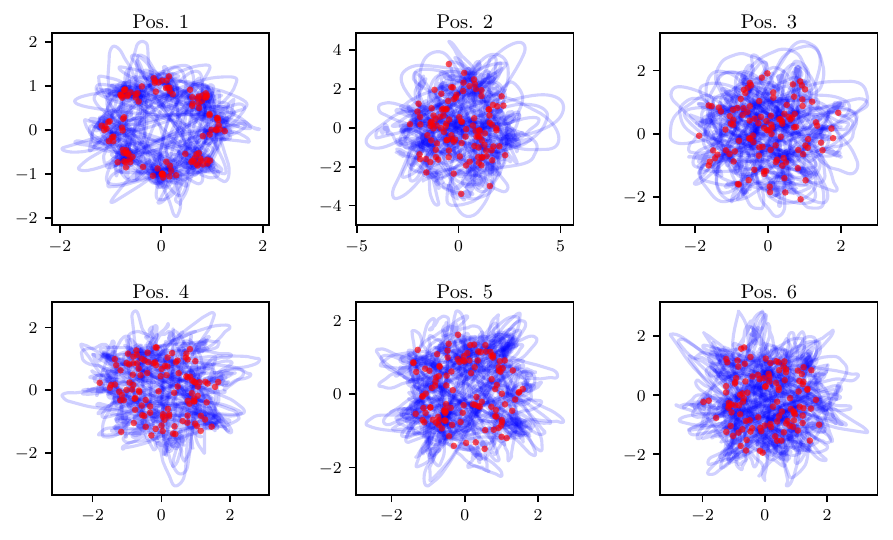}
	\caption{A single recovered constellation at each location for the proposed approach. As the receiver moves off-broadside the constellation becomes more distorted which aids in the security of the transmitted data.}
	\label{recovered_symbols_secure}
\end{figure*}

\begin{figure*}[t!]
	\centering
	\includegraphics[width=0.75\textwidth]{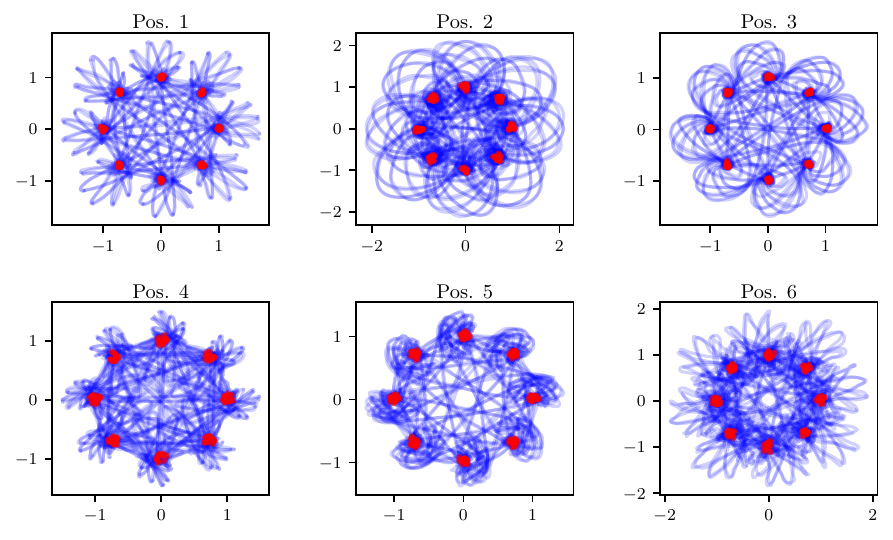}
	\caption{A single recovered constellation at each location for the traditional beamforming approach. With a traditional beamforming approach an eavesdropper can recover the data at all locations measured.}
	\label{recovered_symbols_traditional}
\end{figure*}

\subsection{Experimental Results}
Experiments were conducted for both traditional beamforming, where the same data is sent from each node, along with the secure spatial decomposition approach. 
The measurement results are shown in Table \ref{Table:ser}. The constellation diagrams for a single waveform at each position are shown in Fig.~\ref{recovered_symbols_secure} for the secure approach and Fig.~\ref{recovered_symbols_traditional} for the traditional approach. At broadside both approaches are able to recover the intended data, but the secure approach does have a non-zero, but very low, \gls{ser}. This can could be caused by environmental factors along with coordination degradation; because each transmitting node is only transmitting half the information there may be certain environmental and coordination effects that impact the data which is overcome in the traditional approach by transmitting the same data from two locations.
However, at off-broadside angles the secure approach shows a significant increase in SER, indicating that data recover at these locations would be more challenging. This is demonstrated with symbol errors above 20\% for positions 2 through 6; whereas, for the traditional beamforming approach, the data can be fully decoded at all locations, not just broadside. This demonstrates that our approach increases the security over a traditional distributed beamforming array. Table~\ref{Table:snr} shows the average SNR for measurements at the six positions. The SNR is high at all locations, which demonstrates that all symbol errors are due to the secure transmission approach and not due to low SNR. Note also that the coherent gain of the traditional distributed beamforming approach generally yields higher SNR than the secure approach, as expected since the secure approach does not transmit the same data from the two antennas.

\begin{table}[t!]
\begin{center}
\caption{\label{Table:ser}Symbol Error Ratio of Experiments}
\begin{tabularx}{\columnwidth}{*7{c}}
\toprule
Run & Pos. 1 & Pos. 2 & Pos. 3 & Pos. 4 & Pos 5 & Pos. 6 \\ 
\midrule
Traditional & 0  & 0 & 0 & 0 & 0 & 0\\ 
Secure & 0.0082 & 0.71 & 0.46 & 0.36 & 0.28 & 0.54\\
\bottomrule
\end{tabularx}
\end{center}	
\end{table}

\begin{table}[t!]
\begin{center}
\caption{\label{Table:snr}Signal-To-Noise Ratio (Decibels) of Experiments}
\begin{tabularx}{\columnwidth}{*7{c}}
\toprule
Run & Pos. 1 & Pos. 2 & Pos. 3 & Pos. 4 & Pos 5 & Pos. 6 \\ 
\midrule
Traditional & 37.22  & 34.59 & 37.37 & 36.29 & 36.53 & 30.98\\ 
Secure & 34.99 & 28.65 & 34.56 & 35.04 & 34.29 & 29.42\\
\bottomrule
\end{tabularx}
\end{center}	
\end{table}

\section{Sensitivity Analysis}

\begin{figure}[t!]
	\centering
	\includegraphics[width=0.75\columnwidth]{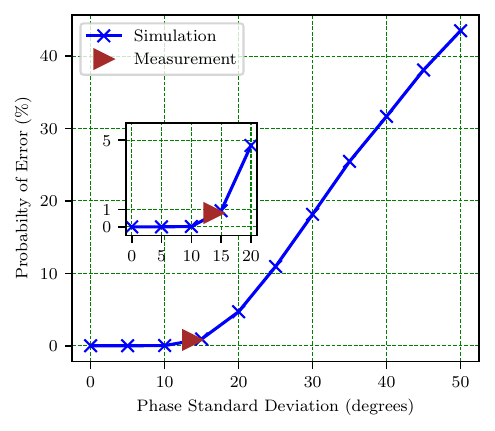}
	\caption{Probability of symbol error compared to varying the standard deviation with zero mean for 8-\gls{psk}.}
	\label{phase_v_error}
\end{figure}

In traditional beamforming and communication approaches, having an SNR above \SI{30}{\decibel} should result in no symbol errors, thus the increase in SER in the secure approach at broadside warrants exploration. 
The principal aspect influencing the errors is likely due to increased phase error between the signals on the secure approach.
 The traditional beamforming approach achieved a phase standard deviation of \SI{7.08}{\degree}. The amplitudes on the up-chirp and down-chirp were nearly identical with amplitudes of \SI{2}{\milli\volt}. For the secure approach, the phase stability between the calibration pulses had a standard deviation of \SI{13.77}{\degree} with amplitudes of \SI{3.2}{\milli\volt} and \SI{2.8}{\milli\volt}.
We implemented a perturbation analysis to determine if the observed SER at broadside is expected with these levels of phase error.
Fig.~\ref{phase_v_error} shows the probability of the data symbol at the receiver being an error as the standard deviation for the phase error increases. 
The probability of error is low if the phase standard deviation is low, and thus it is important to maintain a low phase standard deviation to ensure good performance.
The brown triangle in Fig.~\ref{phase_v_error} is the measurement for the secure case, which had a standard deviation of \SI{13.77}{\degree}. The measured \gls{ser} falls very close to the expected value from the perturbation analysis, thus the errors seen in the measurement at broadside can be attributed to the increase in phase error.

\section{Conclusion}

We demonstrated a novel wireless security approach using distributed coherent transmission and spatial signal decomposition. By separating information between two highly coordinated transmitters, an additional layer of security can be obtained that transmits distorted information to regions away from an intended receiver. Compared to a traditional distributed beamforming approach, the space wherein symbol errors are appreciable is considerably increased. An experimental evaluation demonstrated the feasibility of the approach in a \SI{3}{\giga \hertz} two-element distributed antenna array based on software-defined radios. A perturbation analysis was also implemented to evaluate the impact of relative phase and amplitude errors; this showed that in SER obtained at broadside was due to the measured phase errors in the system, and can also be used determine tolerable error levels in future wireless system designs.

\bibliographystyle{IEEEtran}
\bibliography{dist_comms_reference}

\end{document}